\documentclass[cameraready]{Interspeech}
\usepackage{booktabs}
\usepackage{bbm}
\usepackage{siunitx}
\usepackage{tabularx}
\usepackage{commath}

\usepackage{placeins}
\usepackage{float}
\usepackage{microtype}
\usepackage{tikz}
\usetikzlibrary{positioning, arrows.meta}

\title{Measuring Information Disclosure in Anonymised Speech}

\title{Evaluating Privacy of Speech Anonymisation using Information Theory}

\title{The Equal Error Fallacy: Measuring Information Leakage of Anonymised Speech}

\title{Unequal Errors: Aligning Speech Privacy Metrics with Linkage Threat Models via Calibrated PMI}

\title{Goodbye Equal Error Rate, Hello Evaluating Speech Privacy from the Attacker's Perspective}

\title{Goodbye Equal Error Rate, Hello Local Information Disclosure: An Evaluation of Voice Privacy using Simulated Attacks}

\title{Goodbye Equal Error Rate, Hello Local Information Disclosure: Measuring Speech Anonymisation Privacy using Linkage Attacks}

\title{Goodbye Equal Error Rate, Hello Local Information Disclosure: An Information-Theoretic Approach for the Empirical Evaluation of Voice Anonymisation against 1-to-N Linkage Threats}

\title{Goodbye Equal Error Rate, Hello Local Information Disclosure: Privacy Evaluation of Voice Anonymisation against 1-to-N Linkage Threats}

\title{Goodbye Equal Error Rate, Hello Local Information Disclosure: Evaluating Voice Anonymisation under 1-to-N Linkage Threats}

\title{Goodbye Equal Error Rate, Hello Local Information Disclosure: Evaluating Voice Anonymisation against 1-to-N Linkage Threats}
\author[affiliation={1,2}, orcid=0009-0002-6078-0507]{Dāvis}{Šterns}
\author[affiliation={3}, orcid=0000-0002-3605-7127]{Konstantinos}{Drossos}
\author[affiliation={4}, orcid=0000-0002-9212-7839]{Natasha}{Fernandes}
\author[affiliation={1}, orcid=0000-0002-5590-2349]{Tom}{Bäckström}
\author[affiliation={2}, orcid=0000-0003-4597-7002]{Catuscia}{Palamidessi}

\address{
    $^1$ Aalto University, Finland \\
    $^2$ Inria, France,
    $^3$ Nokia, Finland \\
    $^4$ Macquarie University, Australia
}

\email{davis.sterns@aalto.fi, konstantinos.drosos@nokia.fi, natasha.fernandes@mq.edu.au, tom.backstrom@aalto.fi, catuscia@lix.polytechnique.fr}

\keywords{Voice Anonymisation, Speech Privacy, Empirical Privacy Evaluation, Linkage Attacks}

\usepackage{comment}

\begin{document}

\maketitle

\begin{abstract}
Voice anonymisation aims to protect speaker identity. Currently, its empirical privacy evaluation heavily relies on the Equal Error Rate (EER). Originally designed for biometric verification, EER aggregates scores globally, implicitly assuming an attacker is only trying to verify if two specific voice samples match (a $1$-to-$1$ comparison). This introduces a threat model mismatch with real-world database linkage attacks, where an attacker searches across a fixed set of $N$ enrolled identities (a $1$-to-$N$ closed-set search), allowing global averages to obscure localised privacy failures. While recent $1$-to-$N$ metrics address this aggregation issue, they abstract away the magnitude of the biometric evidence. In this paper, we propose a modular, information-theoretic evaluation framework explicitly designed for the $1$-to-$N$ linkage threat model. 
Within this framework, our core metric, Local Information Disclosure (LID), provides a principled estimate of the identity information disclosed by a single trial utterance in bits by mapping raw similarity scores to an estimated posterior distribution over the enrolled identities. Demonstrating this framework on VoicePrivacy 2024 Challenge systems reveals how global metrics can obscure privacy vulnerabilities. Even for top-performing systems where standard evaluations report near-perfect EERs (48\%), our metric exposes that attackers gain a statistical advantage in at least 63\% of trials with maximum observed disclosures reaching $1$ bit (effectively doubling the attacker's confidence in the target speaker after observing a single trial utterance).
We conclude that shifting toward explainable metrics and appropriate threat models is a practical step toward identifying worst-case vulnerabilities and aligning with strict privacy regulations.
\end{abstract}

\section{Introduction}

Speech signals carry sensitive biometric information, making the widespread collection and processing of voice data a significant privacy risk. To protect the privacy of speakers and to comply with legal frameworks such as the General Data Protection Regulation (GDPR), voice anonymisation has emerged as a method for suppressing speaker identity while preserving linguistic utility~\cite{backstrom2025privacy}. While certain privacy-enhancing technologies offer formal, mathematical guarantees (e.g., differential privacy~\cite{dwork2006calibrating}), extending these theoretical frameworks to the complex, high-dimensional modern generative voice models remains an open challenge. Consequently, current system performance must be evaluated empirically by simulating concrete adversarial attacks against the anonymised data.
\enlargethispage{\baselineskip}

The standard empirical evaluation paradigm, popularised by the VoicePrivacy Challenge (VPC)~\cite{tomashenko2020introducing}, relies on simulating linkage attacks. In this scenario, an attacker computes similarity scores by comparing anonymised \textit{trial} utterances with known identities in an \textit{enrolment} database.\footnote{These terms originate from biometric access control: a user first provides enrolment utterances to register or ``enrol'' their identity in the system, and later provides a ``trial'' sample when attempting to be verified or recognised.} By observing the attacker's performance across numerous trials, auditors can estimate the practical privacy level of the system. To establish a clear threat model, this paper focuses strictly on the risk of linking independent, isolated trial utterances to a known database. We do not address other adversarial objectives, such as inferring attributes like demographic traits~\cite{meyer2023voicepat}, or exploiting longitudinal leakage, where an attacker might leverage temporal patterns across multiple interconnected sessions~\cite{vauquier2025legally}. Under this adversarial framework, privacy evaluation can be described as a two-step process:
\begin{enumerate}
    \item \textbf{Score generation:} Simulating a specific attacker, typically an automatic speaker verification (ASV) system, to compute a matrix of similarity scores between the anonymised trial utterances and the known enrolment profiles.
    \item \textbf{Score evaluation:} Applying a metric\footnote{While a mathematical \textit{metric} formally denotes a pairwise distance function, we adopt the standard colloquial usage of the privacy community, using the term to describe an aggregate statistical measure of system-level risk.} to the resulting similarity score matrix to estimate the overall privacy risk.
\end{enumerate}

While the speech privacy community has actively focused on creating more sophisticated attackers for the score generation phase (as highlighted by the recent VoicePrivacy Attacker Challenge~\cite{tomashenko2025first}), the score evaluation phase still frequently relies on global metrics originally designed for speaker verification~\cite{champion2023anonymizing}. Metrics such as the EER, the Cost of Log-Likelihood Ratio ($C_{\text{llr}}$), Empirical Linkability ($D_{\leftrightarrow}$), and the ZEBRA framework evaluate the statistical separability of isolated mated and non-mated score distributions~\cite{maouche2020comparative}. This reliance introduces a \textit{threat model mismatch} between the two steps: we simulate an attacker who possesses an enrolment database and compares a single trial utterance against \textit{all} enrolments to find the best match (a $1$-to-$N$ search). Yet, by globally pooling these scores into two sets, current metrics evaluate the system as if the attacker were independently guessing whether a single, contextless similarity score belongs to a mated or non-mated pair (a $1$-to-$1$ decision). As we demonstrate in Section~\ref{sec:mismatch}, this mismatch allows standard metrics like the EER to theoretically report a ``safe'' \qty{50}{\percent} error rate even when a simple $1$-to-$N$ linkage attack achieves \qty{100}{\percent} linkage accuracy on the exact same scores.

Alternative score evaluation approaches have recently been proposed to address this aggregation issue, yet to the best of our knowledge, no existing evaluation metric fully aligns with the $1$-to-$N$ threat model simulated in the score generation phase. Similarity Rank Disclosure (SRD)~\cite{backstrom2026privacy} preserves the $1$-to-$N$ relational context, but evaluates the information gain of an attacker who only obtains an ordered list of speakers.
Conversely, the Legally Validated Evaluation Framework~\cite{vauquier2025legally} accounts for probability but relies on the assumption that the attacker makes a forced-choice commitment to a single highest-scoring profile. This highlights an unresolved challenge: quantifying the actual information disclosed to an attacker who observes the complete $1$-to-$N$ vector of similarity scores for a given trial, utilising both the relational context and the magnitude of the evidence.

To address this, we introduce an information-theoretic evaluation framework explicitly designed for the $1$-to-$N$ linkage threat model. The framework takes the complete similarity matrix generated by the attacker and operates in two sequential stages. First, an \textit{inner aggregator} evaluates the information disclosure of each specific trial. This is achieved by calibrating the raw scores into posterior probabilities, and condensing the entire vector of candidate scores for a single trial into one information-theoretic score: the Local Information Disclosure (LID). This captures the strength of the evidence that the attacker gained (or the amount of its confusion). Second, an \textit{outer aggregator} combines these trial-level LID values to construct various global metrics, such as average information disclosure, positive information disclosure rate, and maximum observed disclosure, ensuring that per-trial vulnerabilities remain visible.
The contributions of this paper are as follows:
\begin{itemize}
    \item \textbf{Demonstration of the threat model mismatch:} we examine how existing privacy metrics behave under a $1$-to-$N$ linkage attack. We demonstrate how global metrics obscure local vulnerabilities, and how recent alternative metrics misalign with the attacker's full evidentiary profile, motivating the need for a new approach.
    \item \textbf{Information-theoretic evaluation framework:} we propose a novel privacy metric that provides a calibrated estimate of identity information disclosure in bits. Grounded in pointwise mutual information, our framework measures how much information an attacker gains about a target identity after observing a single trial utterance.
    \item \textbf{Case study on VPC 2024 systems:} To illustrate the framework's utility, we evaluate baselines and top submissions from the VoicePrivacy 2024 Challenge. We demonstrate that our framework enables the detection of localised vulnerabilities in systems with near-perfect EERs.
\end{itemize}

\section{Related Work}
\label{sec:background}
The VPC provides a standard benchmarking framework for empirical evaluation of speech privacy by simulating state-of-the-art ASV attackers. These adversaries typically rely on deep neural architectures, such as ECAPA-TDNN~\cite{desplanques2020ecapa} and x-vectors~\cite{snyder2018x}, to extract speaker embeddings from the audio.
To establish a database of known identities, the embeddings of multiple enrolment utterances are averaged to construct \textit{enrolment profiles}. The adversary then evaluates the similarity between these enrolment profiles and individual \textit{trial} utterances using scoring backends such as probabilistic linear discriminant analysis or cosine similarity.

To systematically evaluate vulnerabilities, the VPC defines a hierarchy of threat models based on attacker knowledge and resources~\cite{tomashenko2024voiceprivacy}. An \textit{ignorant} attacker is unaware of the anonymisation; they use an ASV trained on original speech and compare original enrolment profiles against anonymised trials. A \textit{lazy-informed} attacker is aware of the anonymisation process and aligns the domains by anonymising the enrolment data, but their ASV model remains trained on the original, non-anonymised speech. Finally, the \textit{semi-informed} attacker represents the most serious threat considered: they anonymise the enrolment data and retrain the embedding extractor with the anonymised ASV training corpus to explicitly exploit the residual speaker-discriminative properties of the anonymised domain.

\subsection{Verification under 1-to-1 threat models}
\begin{figure}
    \centering
    \includegraphics[width=1\linewidth]{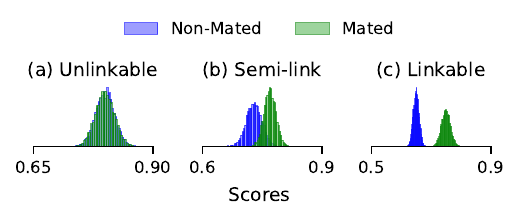}
\caption{Privacy scenarios under the conventional binary evaluation paradigm based on globally pooled score distributions. (a) \textbf{Unlinkable:} The mated (green) and non-mated (blue) distributions overlap completely, preventing speaker discrimination. (b) \textbf{Semi-linkable:} The distributions diverge, creating a ``tail'' of high scores that leaks identity information. (c) \textbf{Fully linkable:} The distributions are separable, allowing for trivial re-identification.}    \label{fig:scenarios}
\end{figure}

In typical evaluation protocols, privacy metrics often frame the attack as a binary hypothesis test. To perform this evaluation, the similarity scores generated across all trials are aggregated into two distinct distributions: a mated set representing same-speaker comparisons, and a non-mated set representing different-speaker comparisons. Privacy is then quantified by assessing the statistical separability of these two sets (as visualised in Figure~\ref{fig:scenarios}). 
To quantify the extent of this overlap, empirical evaluations frequently employ the \textbf{EER}~\cite{tomashenko2024voiceprivacy}.
It is defined as the error rate at the specific decision boundary where the False Acceptance Rate (non-mated comparisons scoring above this boundary) equals the False Rejection Rate (mated comparisons scoring below it).
In an ideal anonymisation scenario with completely overlapping distributions (Figure~\ref{fig:scenarios}a), the EER would be \qty{50}{\percent}, whereas in a fully linkable scenario (Figure~\ref{fig:scenarios}c), it drops to \qty{0}{\percent}.
\textbf{Empirical Linkability} ($D_{\leftrightarrow}$)~\cite{gomez2017general} measures privacy geometrically by calculating the absolute difference between the probability density functions of the two distributions. The \textbf{Cost of Log-Likelihood Ratio} ($C_{\text{llr}}$)~\cite{brummer2006application} applies an information-theoretic approach, evaluating the system based on calibrated log-likelihoods. Finally, the \textbf{ZEBRA framework}~\cite{nautsch2020privacy} extends this probabilistic evaluation by calculating the expected privacy loss across all possible prior knowledge states an attacker might possess. Structurally, these metrics align with a $1$-to-$1$ verification threat model: they model an adversary whose objective is to observe a single similarity score and determine whether it corresponds to a mated or non-mated pair.

\subsection{Identification under 1-to-N threat models}

Recent methodologies adopt a more realistic threat model, recognising that real-world attackers often compare a single intercepted utterance against an entire database of known enrolled identities simultaneously.

\textbf{SRD}~\cite{backstrom2026privacy} evaluates the system by simulating a closed-set linkage task. For each trial utterance, the metric sorts all $N$ available enrolment profiles based on their computed similarity scores. By aggregating these rankings across the evaluation dataset, the method estimates the empirical probability $\gamma_k$ that the target identity appears at rank $k$. The privacy loss is then quantified as the reduction in uncertainty (in bits) relative to a random guess, defined as $\log_2(N \cdot \gamma_k)$. Structurally, SRD models an adversary who observes the ordinal ranking of the candidates, independent of the absolute score values. In practice, this reflects a scenario where an attacker obtains a prioritised shortlist of the most likely matches, ranked from highest to lowest probability.

Bridging empirical technical evaluation with regulatory compliance, \textbf{Legally Validated Evaluation Framework}~\cite{vauquier2025legally} maps GDPR definitions (the risks of Singling Out and Linkability) directly to the ASV outputs. In their framework, empirical linkability is evaluated by calculating the probability that the similarity score of the mated pair is strictly greater than the maximum score among all non-mated pairs in a given trial. This formulation aligns with a threat model in which the adversary is an automated system that must commit to the single highest-scoring enrolment profile as the target identity.

\section{Adversarial Privacy Evaluation Framework}

\label{sec:framework}
In this section, we define the linkage-based adversarial framework that generates the $1$-to-$N$ similarity matrix. A core principle of this framework is that \textit{privacy is not a static, intrinsic property of an anonymisation algorithm}. Instead, privacy is the measurable outcome of the interaction between a protection mechanism, a specific adversary, and the underlying speech dataset. Consequently, swapping the attacker, the protector, or the dataset alters the evaluation outcome. For instance, as demonstrated by Panariello et al.~\cite{panariello2025risks}, evaluating a system against a mismatched attack model creates an illusion of high privacy. This is, however, caused by the failure of the attacker rather than a genuinely robust defence.

\subsection{Threat model specification}

Following the \textit{scenario of use scheme} proposed by Rahman et al.~\cite{rahman2024scenario}, we define the operational boundaries of the evaluation. Table~\ref{tab:threat_model} contrasts the objectives, opportunities, and resources of the attacker and the protector. 

We assume the attacker possesses a database of \textit{enrolment} speech samples mapped to known identities, which are obtained, for example, from public biometric datasets, compromised databases, or social media profiles. Upon intercepting an anonymised \textit{trial} sample, the attacker's objective is to search this database and link the trial sample to the correct identity. Conversely, the protector acts as a data controller who selects and deploys the specific anonymisation method used to protect the trial speech, aiming to prevent this identity linkage while preserving some specific downstream utility of the data.

To accommodate the different threat models defined by the VPC, advanced capabilities are detailed as optional extensions in Table~\ref{tab:threat_model}. Specifically, the ignorant model assumes default access, the lazy-informed model requires knowledge of the deployed anonymisation system, and the semi-informed model demands both this knowledge and the computational resources necessary to train adapted ASV models.

\begin{table*}[htbp]
\centering
\caption{Scenario of Use Specification for Voice Linkage Attacks (adapted from~\cite{rahman2024scenario})}
\label{tab:threat_model}
\renewcommand{\arraystretch}{1.3}
\begin{tabular}{|p{0.1\linewidth}|p{0.4\linewidth}|p{0.4\linewidth}|}
\hline
\textbf{Dimension} & \textbf{Attacker model (attack on privacy)} & \textbf{Protector model (protection of privacy)} \\ \hline
\textbf{Objective} & Successfully link an isolated trial voice sample to a specific identity within a known database. & \textbf{Defence objective:} Prevent speaker re-identification. \newline \textbf{Utility objective:} Ensure the anonymised speech preserves linguistic content and necessary downstream features. \\ \hline
\textbf{Opportunity} & Access to the target speaker's intercepted trial audio. Access to a multi-speaker enrolment database that includes enrolment utterances of the target speaker alongside other distinct identities. \newline \textit{[Optional: Knowledge of the deployed anonymisation system.]} & Full control over the data pipeline to deploy the anonymisation system prior to the audio being intercepted or released. Knowledge that the attacker will employ ASV-based linkage methods. \\ \hline
\textbf{Additional} \newline \textbf{Resources} & Publicly available speech datasets. \newline \textit{[Optional: Finite computing power to train adapted ASV models.]} & Finite computing power. Publicly available speech datasets and baseline ASV models for evaluation. \\ \hline
\end{tabular}

\end{table*}

\subsection{Formal components}

Based on this threat model, we formalise the three interacting components of the evaluation: the dataset, the protector, and the attacker.

\subsubsection{Dataset}

To evaluate speech privacy under this adversarial privacy evaluation framework, we require speech datasets containing multiple speakers, each with multiple utterances. Regardless of the attacker's assumed knowledge level, the auditor's goal is to measure the identity information that can be extracted from the resulting scores. To accurately map the attacker's raw similarity scores into likelihood ratios, the auditor requires two distinct datasets with non-overlapping speaker identities: a development set ($\mathcal{D}_{\text{dev}}$) to calibrate the scoring mechanism, and an evaluation set ($\mathcal{D}_{\text{eval}}$) to compute the final privacy disclosure metrics.

Each of these datasets is further partitioned into an \textit{enrolment} set ($\mathcal{E}$) and a \textit{trial} set ($\mathcal{T}$). 
The enrolment set ($\mathcal{E}$) contains multiple enrolment utterances per speaker, whose embeddings are averaged to form stable identity profiles. The trial set contains distinct utterances from those same speakers, representing the intercepted audio samples.\footnote{While baseline VPC protocols include ``impostor'' trials without enrolled identities to test open-set rejection, our framework explicitly models a closed-set $1$-to-$N$ linkage attack. Thus, we only use trials that possess a target identity in $\mathcal{E}$.}

\subsubsection{Protector}
We define the voice anonymisation mechanism as a function $f$ that transforms an input utterance $x$ into an anonymised utterance $\hat{x} = f(x)$, designed to suppress speaker identity while preserving utility. When applied to a set, $f$ anonymises all utterances in the set.

\subsubsection{Attacker}
We model the attacker as a linkage function $L$ that computes a similarity score $s$ between a trial utterance $\tau_i \in \mathcal{T}$ and a set of enrolment utterances $\varepsilon_j \subset \mathcal{E}$, where $\varepsilon_j$ consists of all utterances of a specific person $j$. Higher values of $s$ correspond to a higher confidence of an identity match. Depending on the attacker's knowledge level (defined in Section~\ref{sec:background}), the inputs to $L$ are modified:
\begin{itemize}
    \item \textbf{Ignorant:} $s = L(\varepsilon_j, f(\tau_i))$.
    \item \textbf{Lazy~/~Semi-Informed:} $s = L(f(\varepsilon_j), f(\tau_i))$ (with the semi-informed $L$ additionally fine-tuned on anonymised data).
\end{itemize}

\subsection{Score generation: the similarity score matrices}
\label{subsec:score_generation}

The attacker simulates a closed-set search independently for $\mathcal{D}_{\text{dev}}$ and $\mathcal{D}_{\text{eval}}$. For each trial utterance $\tau_i \in \mathcal{T}$, the attacker exhaustively queries the linkage function $L$ against all available enrolment profiles in $\mathcal{E}$.

This execution yields two separate similarity matrices: a development matrix $\mathbf{S}_{\text{dev}} \in \mathbb{R}^{T' \times N'}$ and an evaluation matrix $\mathbf{S}_{\text{eval}} \in \mathbb{R}^{T \times N}$, where $T$ and $N$ denote the number of trial utterances and enrolled identities in the evaluation set, respectively, and $T'$ and $N'$ represent the corresponding dimensions for the development set. In both matrices, each row represents the complete evidence for a given trial utterance obtained by the attacker. The generation of such matrices has previously proven effective for visually exposing speaker-level privacy vulnerabilities~\cite{noe2020speech}.

\section{Threat Model Alignment in Privacy Evaluation}
\label{sec:mismatch}

When an evaluation metric's underlying assumptions misalign with the simulated attack, privacy estimates can distort real-world risk. In this section, we illustrate how existing evaluation paradigms misalign with the $1$-to-$N$ linkage threat model.

\subsection{Global aggregation vs. localised search}

Because standard binary metrics assess global separability, they inherently evaluate a $1$-to-$1$ verification threat model. To illustrate why applying this to a $1$-to-$N$ linkage attack misrepresents risk, consider the hypothetical $4 \times 4$ similarity matrix $S$ generated by comparing four trial utterances ($\tau_1$ to $\tau_4$) against four enrolment profiles ($\varepsilon_1$ to $\varepsilon_4$), with mated scores in \textbf{bold}: 

\begin{equation}
\label{eq:score_matrix}
S = 
\begin{bmatrix}
\textbf{0.9} & 0.8 & 0.8 & 0.8 \\
\textbf{0.7} & 0.6 & 0.6 & 0.6 \\
0.4 & \textbf{0.5} & 0.4  & 0.4 \\
0.2 & 0.2 & \textbf{0.3} & 0.2 
\end{bmatrix}.
\end{equation}

A hypothetical attacker operating under a localised $1$-to-$N$ search could simply select the maximum score per row, successfully linking every utterance with \textbf{{\qty{100}{\percent}} accuracy}. 

However, evaluating the matrix with a binary metric yields an \textbf{EER of exactly \qty{50}{\percent}}, a value traditionally interpreted as absolute privacy. This is because the metric evaluates the matrix by stripping away its row-wise context. When aggregated, the set of mated comparisons $\{0.9, 0.7, 0.5, 0.3\}$ and the set of non-mated comparisons $\{ 0.8_{[\times 3]}, 0.6_{[\times 3]}, 0.4_{[\times 3]}, 0.2_{[\times 3]} \}$ overlap. If a decision boundary is drawn at $0.55$, exactly half of the mated comparisons fall below it (a \qty{50}{\percent} False Rejection Rate) and half of the non-mated comparisons fall above it (a \qty{50}{\percent} False Acceptance Rate). 

This vulnerability also extends to any other metric that estimates privacy from the statistical separability of globally pooled score distributions. By treating privacy as a property of an attacker observing a single, contextless similarity score, these metrics rely on global variance in a way that entirely masks the local per-trial leakage.

\subsection{Evidentiary weight in localised search}

While more privacy-centred metrics resolve this global aggregation issue, their specific assumptions about the attacker's behaviour may be misaligned. 

By reducing continuous scores to ordinal ranks, SRD preserves the row-wise context but abstracts away the biometric margin. For instance, SRD evaluates the trial vectors $[\textbf{0.90}, 0.20, 0.10]$ and $[\textbf{0.90}, 0.89, 0.10]$ as mathematically identical Rank-1 observations. Yet, to an attacker analysing these scores, the former scenario yields high confidence due to the significant contrast between scores, whereas the narrow margin in the latter provides weak evidentiary support for a definitive linkage.

Similarly, the Legally Validated Evaluation Framework anchors the localised evaluation in a forced-choice assumption. By strictly measuring the empirical probability that the mated score is the highest, the metric assumes that the attacker must definitively commit to the top-ranked candidate, regardless of the margin. 

In practice, adversaries can leverage the weight of the observed similarity scores. If the scores for several profiles are very similar, the attacker remains uncertain. Conversely, a large margin between the highest score and the rest can increase the attacker’s confidence in the highest-scoring candidate. This motivates a metric that accounts for the complete score vector and the relative evidence it provides across the enrolled candidates.
\section{Proposed Privacy Evaluation Framework}
\label{sec:proposed_metric}

To quantify the information an attacker extracts from the observed similarity scores, we propose a modular, information-theoretic evaluation framework. This framework processes the the two similarity matrices generated in Section~\ref{subsec:score_generation} ($\mathbf{S}_{\text{dev}}$ and $\mathbf{S}_{\text{eval}}$) through three sequential stages: (1) Score Calibration, (2) Inner Aggregation (per-trial local leakage), and (3) Outer Aggregation (global risk profiling).

\subsection{Step 1: Calibration (bridging scores to confidence)}
\label{subsec:calibration}
Raw similarity scores produced by the linkage function $L$ (e.g., cosine similarities) are domain-dependent and lack intrinsic statistical meaning. To quantify the actual evidentiary weight of an attack, we must calibrate these raw scores into probabilistically meaningful Log-Likelihood Ratios (LLRs). We achieve this in two steps: row-wise normalisation followed by logistic calibration.

\subsubsection{Row-wise contextual normalisation.} 
We empirically observe that, particularly for strong anonymisation mechanisms, the absolute magnitude of a similarity score is less informative than its relative distinctiveness among competing enrolment profiles. Therefore, for a given similarity matrix, we first compute the row-wise z-score (often referred to as T-norm in biometrics~\cite{auckenthaler2000score}) for each trial $i$ and enrolment profile $j$:
\begin{equation}
    \hat{s}_{i,j} = \frac{s_{i,j} - \mu_i}{\sigma_i},
\end{equation}
where $\mu_i$ and $\sigma_i$ are the mean and standard deviation of the raw scores across the $N$ enrolled identities for trial $i$.

\subsubsection{Logistic calibration.}
Next, we map these standardised scores to log-posterior odds using a univariate logistic regression model (Platt scaling~\cite{platt1999probabilistic}).
Furthermore, because a $1$-to-$N$ linkage attack suffers from significant class imbalance, where the number of non-mated pairs ($N_{\text{non-mated}}$) in the development set far exceeds the number of mated pairs ($N_{\text{mated}}$) (there is only one mated score per trial), the posterior probabilities learned by the model are naturally skewed. To isolate the pure evidentiary value (the LLR) independently of this prior distribution, we subtract the prior log-odds from the logistic regression output:
\begin{equation}\label{eq:calibration}
    \mathrm{LLR}_{i,j} = (w \hat{s}_{i,j} + b) - \ln\left(\frac{N_{\text{mated}}}{N_{\text{non-mated}}}\right),
\end{equation}
where $w$ and $b$ are the scalar weight and bias learned from fitting the logistic regression on $\mathbf{S}_{\text{dev}}$. Once learned, this calibration pipeline is applied to the evaluation data: $\mathbf{S}_{\text{eval}}$ undergoes the identical row-wise normalisation before the parameters $w$ and $b$ are applied to generate the final evaluation LLRs.

\subsection{Step 2: The inner aggregator (local leakage)}
\label{subsec:inner_aggregator}

The Inner Aggregator estimates the information an attacker extracts from a single observation. For a given trial $i$, let $m_i \in \{1, 2, \dots, N\}$ denote the fixed, ground-truth index of the target identity. However, because this truth is unknown to the attacker, the attacker models the identity of the speaker as a categorical random variable $I \in \{1, 2, \dots, N\}$.

Let $E_i$ represent the biometric evidence observed by the attacker for trial $i$, which in our framework is the vector of calibrated LLRs. By applying the softmax function to these LLRs, we calculate the attacker's posterior probability distribution over all $N$ hypotheses. From this distribution, we extract the specific probability assigned to the correct ground-truth identity ($I = m_i$), given the observed evidence:
\begin{equation}
    P(I = m_i \mid E_i) = p_{i, m_i} = \frac{\exp(\mathrm{LLR}_{i,m_i})}{\sum_{k=1}^N \exp(\mathrm{LLR}_{i,k})}.
\end{equation}
Note that in this equation the bias $b$ and prior log-odds introduced in Equation \ref{eq:calibration} cancel out. Consequently, only the fitted slope $w$ affects the posterior distribution and the resulting scores.

\begin{figure}[t]
    \centering
    \includegraphics[width=1\linewidth]{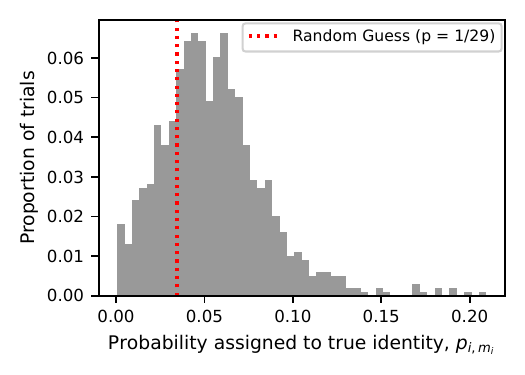}
    \caption{Attacker confidence distribution (T10-2). Empirical distribution of the posterior probability assigned to the target identity ($p_{i, m_i}$) across a dataset of evaluation trials. The red dotted line denotes the baseline prior probability of a random guess ($1/N$, where $N=29$).}
    \label{fig:prob_dist}
\end{figure}

Figure~\ref{fig:prob_dist} visualises the empirical distribution of $p_{i, m_i}$, the probability mass the attacker correctly assigns to the true target across all evaluation trials. While these probabilities are useful for gaining insight into the attacker's performance, to quantify the privacy loss, we can measure the reduction in the attacker's uncertainty.

In information theory, the information gained about one event from observing another is defined by Pointwise Mutual Information (PMI)~\cite{fano1961transmission}. Here, we measure the PMI between the true target hypothesis ($I = m_i$) and the observed evidence ($E_i$):
\begin{equation}
    \mathrm{PMI}(I=m_i ; E_i) = \log_2 \left( \frac{p(I=m_i, E_i)}{P(I=m_i)p(E_i)} \right).
\end{equation}

Estimating the probability density function of observing a specific continuous biometric vector, $p(E_i)$, is infeasible. However, by applying the definition of conditional probability, $P(I, E) = P(I \mid E)p(E)$, the evidence term cancels out:
\begin{equation}
\mathrm{PMI}(I=m_i ; E_i) = \log_2 \left( \frac{P(I=m_i \mid E_i)}{P(I=m_i)} \right).
\end{equation}
This derivation shows that the information gained by the attacker is the log-ratio of their posterior belief to their prior belief. 
A similar conceptual framing of uncertainty reduction is utilised in the SRD framework, which measures the expected information disclosed for each target identity's rank. In contrast, to evaluate the specific evidence observed in a single search, we measure the information disclosed per trial.

Assuming no prior biometric evidence, the attacker's initial belief over enrolled identities is uniform: $P(I=m_i) = 1/N$.
By substituting the prior ($1/N$) and posterior ($p_{i, m_i}$) into the PMI formulation, we yield our proposed metric, \textbf{Local Information Disclosure (LID)}, denoted as $\mathrm{LID}_i$ for trial $i$. The term ``local'' emphasises that this disclosure is measured for each individual trial:
\begin{equation}
    \mathrm{LID}_i = \log_2 \left( \frac{p_{i, m_i}}{1/N} \right) = \log_2(N \cdot p_{i, m_i}).
\end{equation}
This metric captures the system's behaviour after observing a single trial utterance:
\begin{itemize}
    \item \textbf{$\mathrm{LID}_i > 0$ (Positive Disclosure):} The evidence increased the attacker's confidence in the target identity above random guess.
    \item \textbf{$\mathrm{LID}_i = 0$ (No Disclosure):} The prior and posterior match. The evidence provided zero information about the true-identity hypothesis.
    \item \textbf{$\mathrm{LID}_i < 0$ (Active Deception):} The evidence actively misdirected the attacker, lowering their confidence in the target identity to below that of a random guess.
\end{itemize}

To illustrate, consider a trial compared against $N=6$ enrolments yielding raw scores $[0.9, 0.7, 0.4, \mathbf{1.1}, 1.2, 0.4]$ (true target in bold). Row-wise z-normalisation yields $[0.37, -0.27, -1.22, \mathbf{1.01}, 1.33, -1.22]$. Applying calibration (Equation~\ref{eq:calibration}, with $w=1.5, b=-1.0$, and a prior odds ratio of $1/5$) yields the calibrated log-likelihood ratios $[1.17, 0.21, -1.23, \mathbf{2.12}, 2.61, -1.23]$. Then the softmax function converts these LLRs to probabilities $[0.12, 0.05, 0.01, \mathbf{0.31}, 0.50, 0.01]$. Although the true target ($0.31$) is not the top-ranked candidate (reflecting attacker uncertainty), this confidence still exceeds the random-guess prior ($1/6 \approx 0.17$). Consequently, the trial leaks $\mathrm{LID}_i = \log_2(6 \times 0.31) \approx 0.90$ bits of information about the true target speaker.

\begin{figure}[t]
    \centering
    \includegraphics[width=1\linewidth]{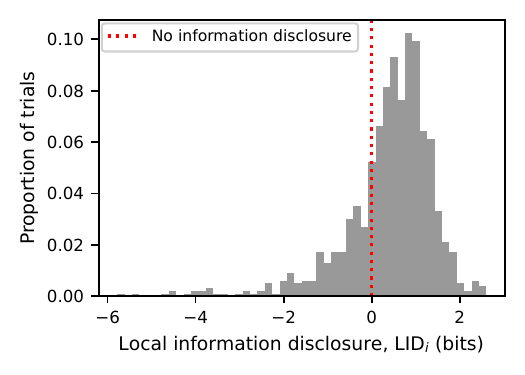}
    \caption{Distribution of $\mathrm{LID}_i$ across evaluation trials (T10-2).}
    \label{fig:infodisc_dist}
\end{figure}

Plotting the distribution of $\mathrm{LID}_i$ can reveal key characteristics of the privacy evaluation. For instance, in the specific example shown in Figure~\ref{fig:infodisc_dist}, while most trials exhibit active privacy leakage ($\mathrm{LID}_i > 0$), the negative tail indicates rare but strong instances of deception.

\subsection{Step 3: The outer aggregator (global risk profile)}
\label{subsec:outer_aggregator}

Once $\mathrm{LID}_i$ is calculated for all $T$ trials in the evaluation dataset, we obtain a global risk vector $\mathcal{L} = [\mathrm{LID}_1, \mathrm{LID}_2, \dots, \mathrm{LID}_T]$. While visual inspection of this distribution provides insight into system dynamics, benchmarking and systematically comparing different anonymisation architectures requires aggregating this vector into scalar global metrics.

The most direct measure of overall privacy loss is the simple average of the trial-level disclosures, which we define as the \textbf{Average Local Information Disclosure (ALID)}:
\begin{equation}
    \mathrm{ALID} = \frac{1}{T}\sum_{i=1}^T \mathrm{LID}_i.
\end{equation}

However, reporting only $\mathrm{ALID}$ can be misleading, as extreme instances of active deception ($\mathrm{LID}_i \ll 0$) can easily cancel out steady positive disclosures ($\mathrm{LID}_i > 0$). To prevent global averages from hiding localised vulnerabilities, we analyse positive and negative disclosures independently.

We define the \textbf{Positive Disclosure Rate} (PDR) as the proportion of trials where the attacker gained a statistical advantage over random chance, calculated as $\mathrm{PDR} = \frac{1}{T}\sum_{i=1}^{T} \mathbbm{1}(\mathrm{LID}_i > 0)$. Conversely, the \textbf{Negative Disclosure Rate} (NDR) is defined as the proportion of trials pushing the attacker into uncertainty or misdirection, computed as $\mathrm{NDR} = 1 - \mathrm{PDR}$. By convention, NDR also includes no-disclosure trials ($\mathrm{LID}_i=0$).

Next, we define the \textbf{Average Positive Disclosure} ($\mathrm{LID}^+$) to strictly measure the expected privacy leakage in bits only during trials where leakage actually occurred: $\mathrm{LID}^+ = \frac{1}{T \cdot \mathrm{PDR}} \sum_{\mathrm{LID}_i > 0} \mathrm{LID}_i$. Similarly, we define the \textbf{Average Negative Disclosure} ($\mathrm{LID}^-$) to measure the average strength of the uncertainty introduced during trials yielding negative disclosure: $\mathrm{LID}^- = \frac{1}{T \cdot \mathrm{NDR}} \sum_{\mathrm{LID}_i \le 0} \mathrm{LID}_i$.

This conditional framework provides a transparent view of the system's behaviour. We can then express the overall average local information disclosure (ALID) as the weighted sum of these two opposing forces:
\begin{equation}
    \mathrm{ALID} = \mathrm{PDR} \cdot \mathrm{LID}^+ + \mathrm{NDR} \cdot \mathrm{LID}^-.\end{equation}
By decomposing the global risk in this manner, auditors can explicitly see both the frequency of positive information disclosure and the precise average leakage within those specific trials, fully isolated from the masking effects of the system's deceptive capabilities.

Finally, even conditional metrics like $\mathrm{LID}^+$ can still mask localised failures. For example,  a system protecting \qty{99}{\percent} of trials perfectly but exposing \qty{1}{\percent} with high confidence still presents a ``singling out'' risk under GDPR. To bound this, we report the \textbf{Maximum Observed Disclosure} ($\mathrm{LID}_{\max} = \max(\mathcal{L})$), which explicitly states the largest information disclosure handed to the attacker across the entire evaluation. The summary of these metrics can be found in Table~\ref{tab:privacy_metrics}.

\begin{table*}[hbtp]
\centering
\caption{Summary of Proposed Privacy Metrics}
\label{tab:privacy_metrics}
\small 
\begin{tabular}{@{}lll p{8.5cm}@{}}
\toprule
\textbf{Metric Name} & \textbf{Symbol} & \textbf{Equation} & \textbf{Description} \\
\midrule
Local Info. Disclosure & $\mathrm{LID}_i$ & $\log_2(N \cdot p_{i, m_i})$ & \footnotesize Bits of evidence leading towards the target speaker obtained by observing the $i$-th trial utterance. \\
\addlinespace
Average LID & $\mathrm{ALID}$ & $\frac{1}{T}\sum_{i=1}^{T} \mathrm{LID}_i$ & \footnotesize The average information disclosure among all trials. \\
\addlinespace
Positive Disclosure Rate & $\mathrm{PDR}$ & $\frac{1}{T}\sum_{i=1}^{T} \mathbbm{1}(\mathrm{LID}_i > 0)$ & \footnotesize Proportion of trials where the attacker gains information. \\
\addlinespace
Negative Disclosure Rate & $\mathrm{NDR}$ & $1 - \mathrm{PDR}$ & \footnotesize Proportion of trials deceiving the attacker. \\
\addlinespace
Average Positive Disc. & $\mathrm{LID}^+$ & $\frac{1}{T \cdot \mathrm{PDR}} \sum\limits_{\mathrm{LID}_i > 0} \mathrm{LID}_i$ & \footnotesize Expected privacy leakage (bits) among the positive disclosure trials. \\
\addlinespace
Average Negative Disc. & $\mathrm{LID}^-$ & $\frac{1}{T \cdot \mathrm{NDR}} \sum\limits_{\mathrm{LID}_i \le 0} \mathrm{LID}_i$ & \footnotesize Average disclosure across non-positive disclosure trials. \\
\addlinespace
Maximum Observed Disc. & $\mathrm{LID}_{\max}$ & $\max(\mathcal{L})$ & \footnotesize Maximum single-trial disclosure. \\
\midrule
\multicolumn{4}{@{}p{\linewidth}@{}}{
\scriptsize \textbf{Notation:} $N$: Number of enrolled identities; $p_{i, m_i}$: Posterior probability of the correct target identity for trial $i$; $T$: Total number of trials in the evaluation dataset; $\mathbbm{1}(\cdot)$: Indicator function yielding 1 if true, 0 otherwise; $\mathcal{L}$: Global risk vector containing all evaluated $\mathrm{LID}_i$ values.
} \\
\bottomrule
\end{tabular}
\end{table*}
\section{Experiments}

\begin{figure*}[t]
    \centering
    \includegraphics[width=0.8\linewidth]{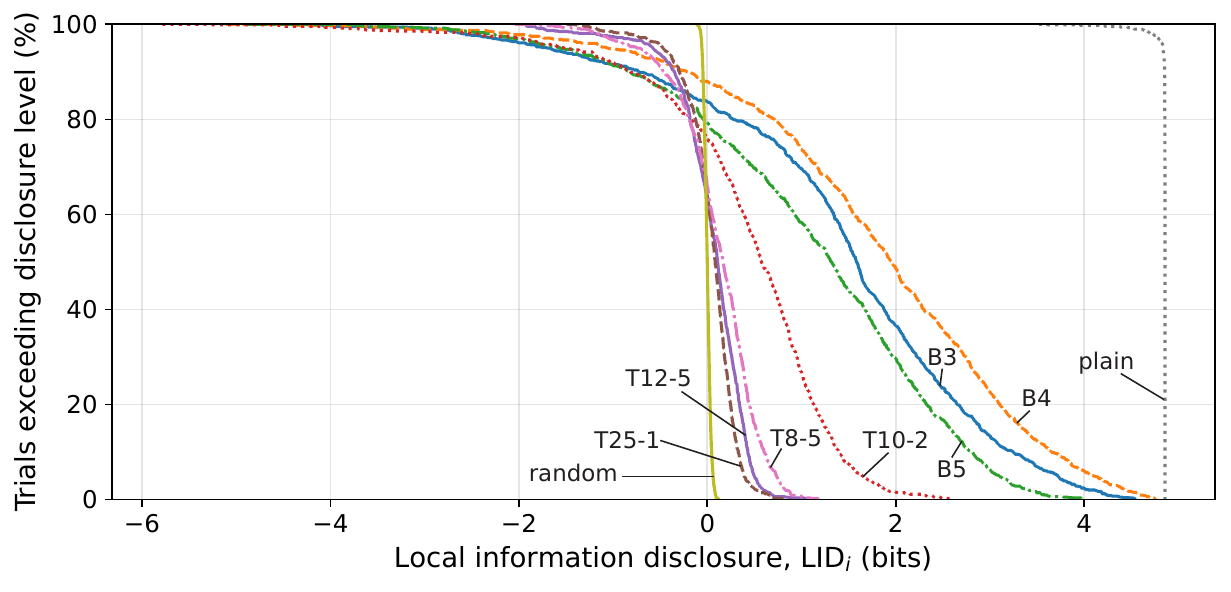}
    \caption{Complementary Cumulative Distribution Function (CCDF) of the localised information disclosure ($\mathrm{LID}_i$) across all evaluation trials. The y-axis denotes the percentage of trials exceeding the privacy disclosure indicated on the x-axis.}
    \label{fig:final_plot}
\end{figure*}

To demonstrate the utility of our framework, we present a case study evaluating a representative selection of anonymisation systems from the VPC 2024. Following the methodology established by \cite{chandra2026evaluating}, we evaluate baselines B3, B4, and B5, alongside systems T8-5, T10-2, T12-5, and T25-1. We simulate a semi-informed attacker using an ECAPA-TDNN architecture on the standard VPC enrol/trial splits of LibriSpeech \textit{dev} and \textit{eval}~\cite{panayotov2015librispeech}. To bound the evaluation, we include a \textit{plain} baseline representing unmodified original speech ($f(x)=x$) and a \textit{random} baseline representing theoretical unlinkability (embeddings generated from independent random noise). To ensure reproducibility, the code used to generate the reported results is publicly available at \url{https://github.com/Speech-Interaction-Technology-Aalto-U/LID}.

Following the procedure defined in Section~\ref{subsec:calibration}, we apply row-wise z-normalisation and fit a univariate logistic regression model to the \textit{dev} matrix to predict ground-truth labels. The learned calibration weights $w$ and biases $b$ are detailed in Table~\ref{tab:wb_metrics}.

The weight parameter $w$ serves as a direct indicator of the predictive power of the underlying raw scores, dictating how aggressively the log-likelihood scales as a normalised score deviates from the trial's mean. The \textit{plain} baseline yields a highly predictive weight ($w=3.06$), indicating that a score's relative distinctiveness within a trial strongly correlates with the target identity. Conversely, the \textit{random} baseline correctly collapses toward zero ($w=0.03$).

\begin{table}[htbp]
\centering
\caption{Logistic regression calibration parameters learned on the disjoint \textit{dev} dataset.}
\begin{tabular}{lrr}
\toprule
\textbf{Experiment} & $\mathbf{w}$ & $\mathbf{b}$  \\
\midrule
B3      & 1.53 & -0.99 \\
B4      & 1.86 & -1.25 \\
B5      & 1.15 & -0.60 \\
T8-5    & 0.34 & -0.05 \\
T10-2   & 0.92 & -0.32 \\
T12-5   & 0.30 & -0.04 \\
T25-1   & 0.23 & -0.02 \\
\midrule
plain   & 3.06 & -5.56 \\
random  & 0.03 &  0.00 \\
\bottomrule
\end{tabular}
\label{tab:wb_metrics}
\end{table}

These parameters are used to calibrate the \textit{eval} similarity matrix, and the Inner Aggregator calculates the localised information disclosure ($\mathrm{LID}_i$) for every individual trial. Figure~\ref{fig:final_plot} visualises the Complementary Cumulative Distribution Function (CCDF) of these trial-level disclosures. This representation systematically unmasks the local behaviour of each anonymisation method. The extreme cases are showcased by the \textit{plain} baseline (where every single trial leaks more than $3.5$ bits of information) and the \textit{random} baseline (exhibiting a near-vertical collapse at $0$ bits, with only marginal deviations due to statistical noise). By mapping the exact proportion of trials at every disclosure level, this visualisation allows auditors to instantly quantify the density of severe localised privacy failures.

\begin{table*}[htbp]
\centering
\caption{Comparison of reference metrics (EER, $C_{\text{llr}}$) and average Rank Disclosure (MeanD, sourced from~\cite{chandra2026evaluating}) against our proposed $1$-to-$N$ metrics: ALID, Positive/Negative Disclosure Rates (PDR/NDR), conditional average disclosures ($\mathrm{LID}^+$/ $\mathrm{LID}^-$), and the maximum observed disclosure ($\mathrm{LID}_{\max}$). Arrows ($\uparrow$/$\downarrow$) indicate the direction of increased privacy.}
\label{tab:experiment_metrics}
\resizebox{\linewidth}{!}{%
\begin{tabular}{lccccccccc}
\toprule
\textbf{Experiment} & \textbf{EER} ($\uparrow$) & $\mathbf{C}_{\text{llr}}$ ($\uparrow$) & \textbf{MeanD} ($\downarrow$) & \textbf{ALID} ($\downarrow$) & \textbf{PDR} ($\downarrow$) & \textbf{NDR} ($\uparrow$) & $\mathbf{LID}^+$ ($\downarrow$) & $\mathbf{LID}^-$ ($\downarrow$) & $\mathbf{LID}_{\text{max}}$ ($\downarrow$) \\
& & & \textbf{(bits)} & \textbf{(bits)} & \textbf{(\%)} & \textbf{(\%)} & \textbf{(bits)} & \textbf{(bits)} & \textbf{(bits)} \\
\midrule
B3      & 0.24 & 0.72 & 0.06 & 1.41 & 84.0  & 16.0 & 1.94 & \textbf{-1.30} & 4.54 \\
B4      & 0.27 & 0.76 & \textbf{0.02} & 1.80 & 88.0  & 12.0 & 2.21 & -1.17 & 4.75 \\
B5      & 0.28 & 0.78 & \textbf{0.02} & 1.11 & 79.0  & 21.0 & 1.67 & -1.04 & 3.98 \\
T8-5    & 0.45 & 0.98 & 0.06 & 0.10 & 66.0  & 34.0 & 0.35 & -0.38 & 1.18 \\
T10-2   & 0.37 & 0.92 & 0.52 & 0.40 & 76.0  & 24.0 & 0.84 & -0.98 & 2.60 \\
T12-5   & 0.47 & \textbf{0.99} & \textbf{0.02} & \textbf{0.04} & \textbf{63.0}  & \textbf{37.0} & 0.26 & -0.33 & 1.06 \\
T25-1   & \textbf{0.48} & \textbf{0.99} & 0.05 & \textbf{0.04} & 64.0  & 36.0 & \textbf{0.20} & -0.25 & \textbf{0.83} \\
\midrule
plain   & 0.00 & 0.01 & 2.19 & 4.85 & 100.0 & 0.0  & 4.85 & -     & 4.86 \\
random  & 0.50 & 1.00 & 0.00 & 0.00 & 50.0  & 50.0 & 0.03 & -0.03 & 0.12 \\
\bottomrule
\end{tabular}%
    }
\end{table*}

\begin{figure}[htbp]
    \centering
    \includegraphics[width=1\linewidth]{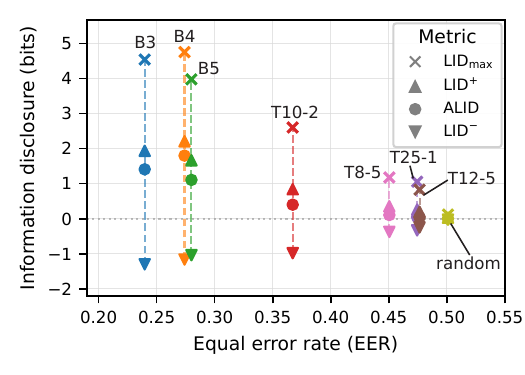}
    \caption{Comparison of Equal Error Rate (EER) against proposed information-theoretic metrics. The divergence of $\mathrm{LID}_{\max}$ highlights that systems with equivalent global EERs can have different observed maximum disclosures.}
    \label{fig:eer_vs}
\end{figure}

The scalar global risk profiles generated by the Outer Aggregator are summarised in Table~\ref{tab:experiment_metrics}. Because our evaluation exhaustively compares all trials against all enrolments, the reported EERs differ from official VPC results. By isolating the Positive Disclosure Rate (PDR) from the Negative Disclosure Rate (NDR), the framework quantifies the precise directionality of the privacy loss. All evaluated systems (excluding the random baseline) exhibit a PDR of at least \qty{63}{\percent}. We observe a positive correlation between the calibration weight $w$ and the magnitude of the conditional disclosures ($\mathrm{LID}^+$ and $|\mathrm{LID}^-|$). A higher $w$ indicates that the system's output distributions possess higher variance in biometric confidence. Consequently, systems like B4 not only yield higher expected leakage during positive disclosure trials ($\mathrm{LID}^+ = 2.21$ bits) but also generate more convincing false evidence during misdirections ($\mathrm{LID}^- = -1.17$ bits). Expanding upon existing metrics like SRD, this continuous approach allows auditors to measure exactly how much statistical confidence the attacker gains alongside the relative positional ranking.

Finally, Figure~\ref{fig:eer_vs} illustrates the additional distinctions revealed by the proposed $1$-to-$N$ metrics. B4 and B5 obtain similar EERs of approximately $0.27$ and $0.28$, respectively, yet B4 exhibits consistently greater disclosure, with higher ALID, $\mathrm{LID}^{+}$, and $\mathrm{LID}_{\max}$. This comparison shows that systems with similar global verification performance can present different local disclosure profiles under the closed-set linkage threat model. Moreover, T12-5 and T25-1 report EERs of 0.47 and 0.48, respectively (values conventionally interpreted as indicating strong anonymisation), yet still exhibit positive disclosure in at least \qty{63}{\percent} of trials and observed maximum disclosures of 1.06 and 0.83 bits. Together, these comparisons show that global verification performance does not fully characterise local disclosure under the $1$-to-$N$ threat model.

\section{Discussion}
\label{sec:discussion}

As illustrated by our theoretical example, global aggregation can yield a \qty{50}{\percent} EER even when an attacker performing a $1$-to-$N$ search correctly re-identifies every trial. This shows that EER does not necessarily reveal trial-level linkage vulnerabilities. Our case study demonstrates the practical relevance of this distinction. B4 and B5 obtain similar EERs of \qty{27}{\percent} and \qty{28}{\percent}, respectively, yet B4 exhibits greater average and maximum observed disclosure than B5, with an ALID of 1.80 versus 1.11 bits and a $\mathrm{LID}_{\max}$ of 4.75 versus 3.98 bits. Thus, similar binary verification performance can coexist with different local disclosure profiles. Likewise, T12-5 and T25-1 obtain EERs of \qty{47}{\percent} and \qty{48}{\percent}, yet their maximum observed disclosures reach 1.06 and 0.83 bits, respectively. Because evaluation metrics guide both system optimisation and deployment decisions, this distinction has practical consequences: optimising EER alone may leave localised linkage disclosure unaddressed.

Binary metrics, such as EER, remain useful for evaluating $1$-to-$1$ speaker verification, but privacy assessment must be tied to specified adversarial behaviour. LID combines two complementary ideas from existing privacy metrics. Following ZEBRA, it uses calibrated evidence to express disclosure in information-theoretic terms. Following SRD and the Legally Validated Evaluation Framework, it evaluates each trial within its complete $1$-to-$N$ candidate context. Unlike ZEBRA's pairwise quantities and rank-based metrics that abstract away score magnitudes, LID uses the complete score vector to estimate the magnitude and direction of the change in posterior support for the true identity. Retaining one disclosure value per trial also exposes variation hidden by a single aggregate. Average disclosure alone may conceal rare high-disclosure trials, whereas the observed maximum alone cannot show whether such disclosure is isolated or widespread. The scalar summaries and CCDF therefore provide complementary views of the trial-level disclosure distribution.

Validating a new privacy metric presents a particular epistemological challenge: there is no definitive meta-metric or absolute ground truth to measure the quality of a privacy evaluation. Consequently, a proposed metric cannot simply be proven ``correct'' or definitively better than other metrics through empirical outperformance. Instead, its validity can be supported through structural alignment with a specified threat model, clear interpretability, and predictable behaviour under controlled reference conditions. 
LID addresses the threat model by mirroring the inputs an attacker observes in a closed-set search.
It addresses interpretability by outputting a clear unit: bits of information.
LID, based on pointwise mutual information, estimates the information gained about the true-identity hypothesis after observing a single trial utterance. Positive values indicate increased posterior support for the true identity, whereas negative values indicate decreased support. This trial-level interpretation is closely related to singling-out risk, because it describes whether an utterance makes its speaker more linkable to the corresponding enrolled identity.
Finally, the inclusion of our \textit{random} and \textit{plain} baselines serves as a necessary sanity check. As demonstrated in our experiments, the average disclosure converges near zero bits for randomly generated speaker embeddings, while the unanonymised speech approaches the theoretical maximum of $\log_2N$ bits. Although reaching these reference points does not establish the accuracy of the underlying probability estimates for every trial, the observed behaviour supports the theoretical interpretation of LID within the evaluated configuration.

Despite these advantages, the proposed framework (like any empirical privacy evaluation) measures information leakage based on a specific interaction between a protector, an attacker, a dataset, and its split. It does not provide static privacy guarantees for a given anonymisation algorithm. A broader system audit would therefore need to consider multiple datasets, galleries, and attacker models. Accordingly, \(\mathrm{LID}_{\max}\) should be interpreted as the single largest disclosure observed during a specific test, not a universal worst-case bound; to fully understand a system's leakage profile, we recommend analysing the CCDF graph alongside scalar metrics. Furthermore, because LID relies on mapping raw similarity scores to log-likelihood ratios, the accuracy of the estimated posterior probabilities remains sensitive to the chosen calibration method. Future work should explore whether alternative calibration techniques provide better-calibrated posterior estimates of an attacker’s true confidence. Extending the framework to longitudinal and multimodal attacks will additionally require methods that account for dependence between multiple observations. 
To conclude, by making localised linkage vulnerabilities visible and quantifiable, this work supports the development and selection of voice anonymisation systems that better protect individuals whose privacy risks might otherwise remain hidden by global averages.

\section{Acknowledgments}
We would like to thank the VPC organizers for sharing the anonymised LibriSpeech datasets, and the Audio Security and Privacy Lab at EURECOM, France, for sharing the ECAPA-TDNN model weights.

This work was funded by the European Union’s Horizon Europe research and innovation programme under grant agreement No. 101168193.

\FloatBarrier
\bibliographystyle{IEEEtran}
\bibliography{my_references}

\end{document}